\documentclass{article}

\usepackage[dblblindworkshop, final]{neurips_2025}

\workshoptitle{ML4PS}

\usepackage[utf8]{inputenc} % allow utf-8 input
\usepackage[T1]{fontenc}    % use 8-bit T1 fonts
\usepackage[colorlinks=true,citecolor=blue,linkcolor=blue,urlcolor=blue, backref=false,pdfborder={0 0 0}]{hyperref}
\usepackage{url}            % simple URL typesetting
\usepackage{booktabs}       % professional-quality tables
\usepackage{amsfonts}       % blackboard math symbols
\usepackage{nicefrac}       % compact symbols for 1/2, etc.
\usepackage{microtype}      % microtypography
\usepackage[table,svgnames,dvipsnames]{xcolor}

\usepackage{amsmath}
\usepackage{graphicx}
\usepackage{sidecap}

\usepackage{orcidlink}

\title{Reconstructing the local density field with combined convolutional and point cloud architecture}

\author{%
  Baptiste Barthe-Gold \\
  Ecole Polytechnique \\
  Rte de Saclay, 91120 Palaiseau, France \\
  \texttt{baptiste.barthe-gold@polytechnique.edu} \\
  \And
  Nhat-Minh Nguyen \orcidlink{0000-0002-2542-7233} \\
  Center for Data-Driven Discovery \\
  Kavli IPMU, UTIAS, The University of Tokyo \\
  Kashiwa, Chiba 277-8583, Japan \\
  \texttt{nhat.minh.nguyen@ipmu.jp} \\
  \And
  Leander Thiele \orcidlink{0000-0003-2911-9163} \\
  Center for Data-Driven Discovery \\
  Kavli IPMU, UTIAS, The University of Tokyo \\
  Kashiwa, Chiba 277-8583, Japan \\
  \texttt{leander.thiele@ipmu.jp} \\
}

\begin{document}

\maketitle

\begin{abstract}
We construct a neural network to perform regression on the local dark-matter density field given line-of-sight peculiar velocities of dark-matter halos, biased tracers of the dark matter field.
Our architecture combines a convolutional U-Net with a point-cloud DeepSets.
This combination enables efficient use of small-scale information and improves reconstruction quality relative to a U-Net-only approach. Specifically, our hybrid network recovers both clustering amplitudes and phases better than the U-Net on small scales.
\end{abstract}

\section{Introduction}

One key objective in observational cosmology is the inference of the local matter density field in the late-time universe.
Given that the matter distribution is dominated by the invisible dark matter, inference must rely on indirect probes.
One such probe is galaxy peculiar velocity. Peculiar velocity is the velocity component of galaxies induced by large-scale matter clustering (separated from the smooth Hubble flow component due to cosmic expansion).

Unlike the formation and clustering of galaxies, the peculiar velocities of galaxies are directly sourced by gravitational interactions of the matter distribution itself, hence a fundamentally unbiased tracer of the matter density field in the local universe, out to the distance of a few hundred Megaparsecs (Mpc), though only the radial, i.e. line-of-sight, component of galaxy peculiar velocities can be measured from their Doppler redshift.

Peculiar velocity determinations rely on direct distance measurements, for example via the fundamental plane, Tully-Fisher relations or supernovae Ia standard candles.
Reconstruction of the local density field from peculiar velocities has a venerable history.
Traditionally, linear methods have been employed, i.e.\ relations of the form $\delta = M v$, where $\delta$ is the (over)density field $\rho/\bar\rho-1$, $v$ is the array of line-of-sight velocities, and $M$ is a linear operator.
The direct inversion technique is an application of linear theory (written in Fourier conjugate $\mathbf{k}$-space and restricted to redshift $z=0$ for simplicity):
\begin{equation}
  \delta_\text{L}(\mathbf{k}) = -\frac{i}{Hf}\mathbf{k}.\mathbf{v}_{\mathrm{LOS}}(\mathbf{k})\,,
  \label{eq:direct}
\end{equation}
where $H$ is the Hubble parameter and $f$ is the linear growth rate of the density field $\delta$, while $\mathbf{v}_{\mathrm{LOS}}$ denotes the line-of-sight velocity vector.
In the context of Bayesian inference, assuming a Gaussian prior on $\delta$, one obtains the alternative linear technique, namely Wiener filtering~\citep{1999ApJ...520..413Z,2012ApJ...744...43C,2024MNRAS.527.3788H}.

Linear reconstruction is necessarily limited to large scales (the regime where $\delta\ll1$) and thus cannot optimally use the dense information available in modern peculiar velocity data.
In order to improve the reconstruction on nonlinear scales, sampling of initial conditions through a forward model has been developed~\citep[e.g.][]{2014ApJ...794...94W,2019A&A...625A..64J,2023MNRAS.518.4191P}.
The computational expense of running a gravity solver in each sampling step and the approximations in the solver limit forward modeling in resolution.
Furthermore, being able to run only a few chains within reasonable computational budgets makes it difficult to explore systematic effects~\citep{2021JCAP...03..058N}.

Therefore, machine learning has been introduced as a possible improvement. 
Recent works use a convolutional neural network to perform the mapping from input data to the density field. Some methods only use the observed (galaxy) density field as input, while others also include the gridded line-of-sight velocities~\citep{Ganeshaiah_Veena_2023,2021ApJ...913...76H}.

Our contribution aims to assess improvements on this method in two aspects.
First, directly passing the gridded radial velocity appears suboptimal given the inductive bias of convolutional networks.
Indeed, except for boundary effects the point of the convolutional network is to exploit translational invariance for weight sharing.
In order to improve the match between inductive bias and problem formulation, we pass the linear reconstruction $\delta_\text{L}$ from Eq.~\eqref{eq:direct}.
Second, working solely with a grid misses small-scale information, particularly in high-density regions. In our convolutional setup, any vorticity information is also neglected.
More involved models have been developed to improve small-scale behavior, such as using graph neural networks along with a learned assignment scheme \cite{kvasiuk2024reconstructioncontinuouscosmologicalfields}.
We propose another method and add a second component to the architecture that is responsible for recovering the small-scale features.
This component is chosen as a point-cloud architecture such that it can be evaluated directly on the local set of surrounding galaxies.

We restrict ourselves to application in numerical simulations in this work.
Evaluation on observational data such as Cosmicflows-4~\citep{2023ApJ...944...94T} requires careful treatment of complicated systematic effects in the training data.
Furthermore, for such application we will need to include stochasticity in the network architechture (presumably in a generative fashion).
For simplicity, we only consider regression of the mean in this work, noting that the inclusion of stochasticity should be relatively straightforward.
In addition, we attempt to prevent information about the tracer density field from informing the prediction, in order to isolate and focus on the information content in peculiar velocities.

After introducing the network and training method in Sec.~\ref{sec:method}, we will show in Sec.~\ref{sec:results} that the neural network improves reconstruction quality on nonlinear scales, i.e. $k=0.1-1\,h\text{Mpc}^{-1}$.

\section{Method}
\label{sec:method}

\subsection{Data}

We train on the high-resolution version of the \texttt{Quijote} fiducial simulations~\citep{2020ApJS..250....2V} at redshift $z=0$.
The single-redshift approximation is highly accurate and sufficient for the typical size of peculiar velocity surveys (see above).

In this work, we assume that the background cosmology is known and fixed, hence the same cosmology for all simulations. Note that this is a standard assumption in most (if not all) reconstructions~\citep[e.g.][]{2016MNRAS.457..172L,2019A&A...625A..64J,2019arXiv190906396L} and their derived astrophysics applications~\cite[e.g.][]{2020JCAP...12..011N,2022JCAP...08..003T,2022PhRvD.106j3526B}. In future works, especially for a joint inference of cosmological parameters and the matter density field, we will additionally consider simulations with varying cosmology.

We further approximate galaxies as dark matter halos, identified with \texttt{Rockstar}~\citep{2013ApJ...762..109B}, a friends-of-friend halo-finding algorithm in 6D phase-space.
For each of the 100 simulation boxes with side length $1\,h^{-1}\text{Gpc}$,
we randomly pick a certain halo as the observer's host halo, then proceed to collect all neighboring halos within a sphere of $200\,h^{-1}\text{Mpc}$ radius.
This radius roughly corresponds to the maximum range of current peculiar velocity surveys.
We thus use $\sim 30,000$ halos per training sample, and generate $256$ of those training samples per simulation, which results in a dataset of $25,600$ items.
We split those into training, validation, and test sets in an 80/10/10 proportion among the 100 \texttt{Quijote} simulations, thereby preventing information leakage.

Constructing the peculiar velocity field from halo peculiar velocities requires first constructing their momentum grid and then dividing by their density grid. Care usually must be taken during the latter to avoid spurious bias due to the sparsity of the halos. The resulting peculiar velocity field is, in any case, inevitably biased low in amplitude. To this end, one might attempt to correct for such bias analytically; here, however, we take a pragmatic approach and smooth the momentum and density fields with a Gaussian kernel after assignment.
For small values of the kernel size, the direct inversion $\delta_\text{L}$ shows large artifacts.
On the other hand, a large kernel size leads to loss of small-scale information.
We thus stack three choices of kernel size ($1, 2, 4\,h^{-1}\text{Mpc}$) as separate channels in the input to the convolutional network and let the learning algorithm find the optimal weighting.
We use a $128^3$ regular grid, corresponding to a cell spacing of $\approx 3.23\,h^{-1}\text{Mpc}$.

\subsection{Network Architecture}

\begin{figure}
\centering
\includegraphics[width=0.8\linewidth]{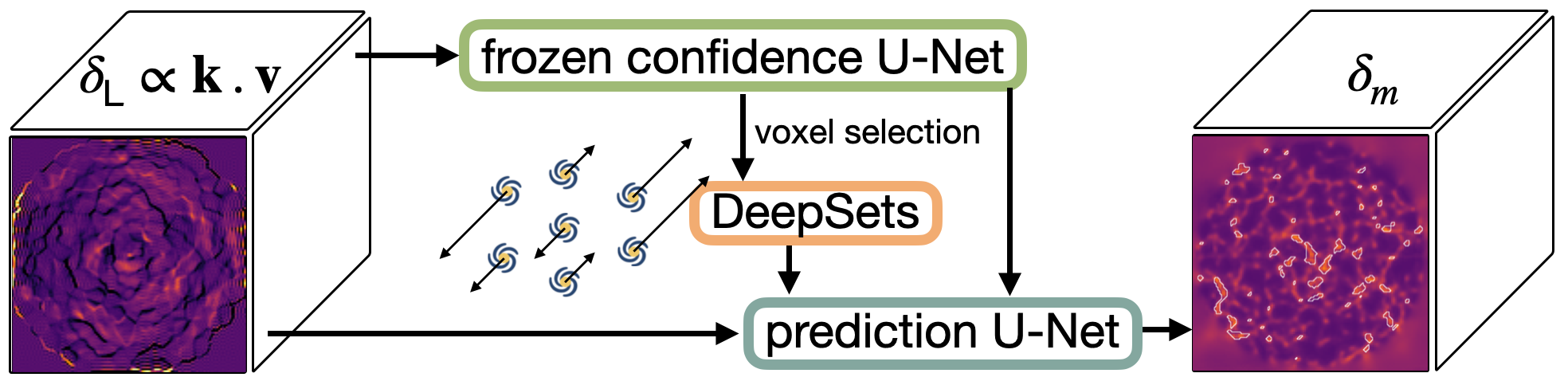}
\caption{Schematic representation of our architecture. The backbone is a U-Net evaluated on the velocity divergence. Additional small-scale information is provided by a DeepSets point-cloud architecture which is evaluated locally. Due to the high cost of the DeepSets evaluation, a pretrained and then frozen confidence U-Net is used to select a small percentage of voxels for which the DeepSets evaluation is deemed worth the expense.}
\label{fig:arch}
\end{figure}

Our architecture is a combination of convolutional and point cloud, as we illustrate schematically in Fig.~\ref{fig:arch}.

For the point-cloud part, we use a DeepSets architecture~\citep{zaheer2018deepsets} which operates on the halos in a $10\,h^{-1}\text{Mpc}$ radius around a given voxel.
The DeepSets uses the relative position and line-of-sight velocity of the halos as point-level data while the global position of the respective voxel and mean line-of-sight velocity in the set are treated as global features.

Computationally, it is challenging to evaluate the DeepSets across the entire volume (it would mean $128^3$ separate DeepSets computations for each training sample).
We thus introduce a trick to make the problem manageable.
We train a separate two-headed U-Net (the ``confidence network'') to predict both the expectation value $\bar\delta$ and the error $\sigma$ in the style of a moment network~\citep{2020arXiv201105991J}.
The confidence network is trained with a v loss~\citep{seitzer2022on} with $\beta=0.5$.
Only for the $N_\text{DS}=5\times10^4$ voxels (about 2.4\% of volume) with the highest predicted error $\sigma$ do we evaluate the DeepSets.
Empirically, the convergence in $N_\text{DS}$ is quite good and we do not expect substantial improvement from assigning more resources to the DeepSets evaluation.

For the convolutional part, we use a U-Net~\citep{10.1007/978-3-319-24574-4_28}
with LeakyReLU activations and layer normalization.
The input channels consist of:
\begin{itemize}
\item the direct inversion linear density field $\delta_\text{L}$ for three different smoothing scales (as described above),
\item the position of the voxel relative to the observer,
\item the DeepSets output (four channels in our setup), and
\item the frozen confidence network output $\bar\delta$.
\end{itemize}
The output $y$ of the U-Net is transformed as $\hat\delta=\text{ReLU}(\sinh(y))-1$ in order to better match the dynamic range of the density field: many values of order unity, but much larger in high-density regions.
A similar transformation was found beneficial in Ref.~\cite{2020ApJ...902..129T}.

\subsection{Network Training}
The training procedure is split into two stages, first training the confidence U-Net model by itself and then training the full model using the confidence U-Net as one of the input channels. The main drawback of this method is having to train two models separately.

We optimize the field-level mean-squared error (or $\beta$-NLL loss for the confidence model) using the AdamW optimizer~\citep{loshchilov2018fixing} with a one-cycle learning rate scheduler~\citep{2017arXiv170807120S}.

The model is trained using 24 Nvidia A100 GPUs. Each model is trained until a convergence of the validation loss, usually reached at around 200 epochs. The training time for the whole model is on the order of 30 hours.

\begin{table}
\centering
\caption{The Mean Squared Error (MSE) obtained by different models evaluated on the test set.}
\begin{tabular}{l c}
\hline
\textbf{Model} & \textbf{MSE} \\
\hline
Direct inversion ($1\,h^{-1}\text{Mpc}$ smoothing) & 4.94 \\
Direct inversion ($4\,h^{-1}\text{Mpc}$ smoothing) & 4.48 \\
3D Wiener filter & 3.80 \\
Normal U-Net & 3.20 \\
Confidence U-Net ($\mu$ prediction) & 3.26 \\
U-Net + DeepSets (50k voxels) & 2.99 \\
\hline
\end{tabular}
\label{tab:mse_comparison}
\end{table}

Table~\ref{tab:mse_comparison} reports the MSE for all models considered, when evaluated on the test set. Most models yield quantitative gains over the baseline U-Net. The only exception is the confidence U-Net although not really unexpected, given its $\beta$-NLL objective does not optimize the MSE. Note that the confidence U-Net only serves as one input channel for the final DeepSets-augmented model, which is trained separately and achieves the lowest MSE among all methods.

\section{Results}
\label{sec:results}

\begin{SCfigure}
\includegraphics[width=0.6\linewidth]{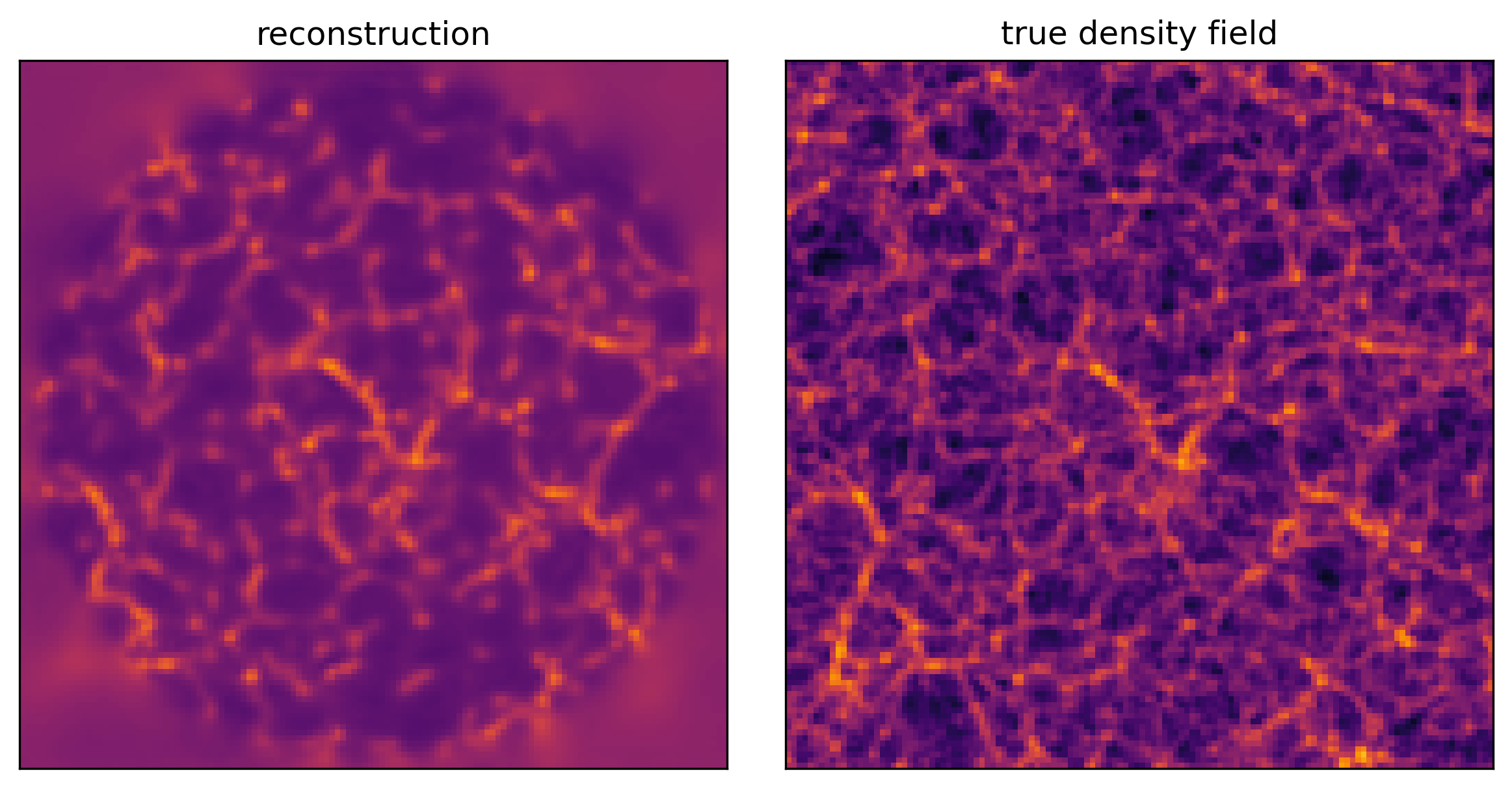}
\caption{The same density field slice through the reconstruction (left) and the truth (right), on logarithmic color scale.}
\label{fig:pic1}
\end{SCfigure}

We begin by presenting our model output in an example from the test set, displayed in Fig.~\ref{fig:pic1}.
The circular geometry in the reconstructed field is due to the selection of tracer halos within a $200\,h^{-1}\text{Mpc}$ radius.
We observe good agreement between the reconstruction and the underlying true density field over a range of scales.
Of course, the reconstructed field loses fidelity on small scales due to the finite tracer density and lack of transverse modes.
Outside the region covered by tracers, the model mostly falls back to the mean, although thanks to the learned cosmic web prior it does predict some vague structures like the density in the lower left.

\begin{figure}
\centering
\includegraphics[width=0.9\linewidth]{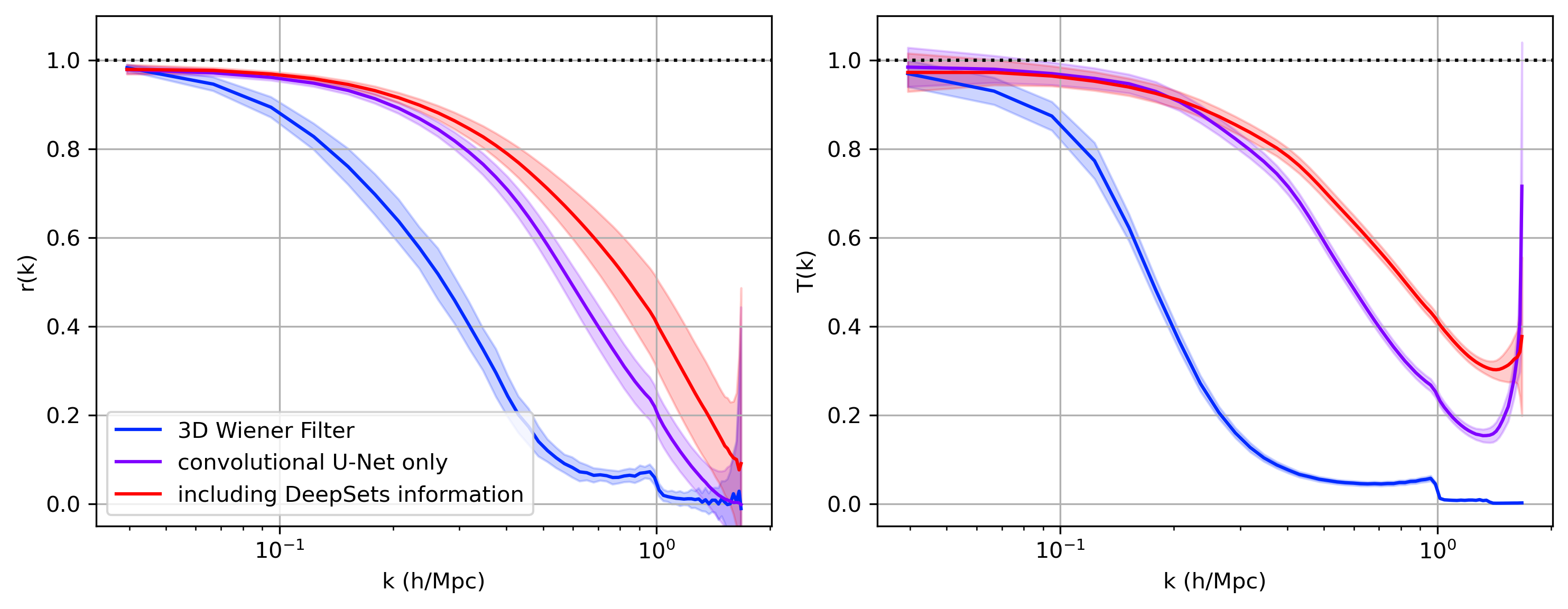}
\caption{Performance evaluation in two-point statistics. The left panel shows the cross-correlation coefficient, while the right panel shows the transfer function. Blue is the 3D Wiener filter (which is an upper bound on linear reconstruction), magenta shows the reconstruction with a convolutional U-Net only, and red shows our complete architecture which includes small-scale information from the DeepSets component. The shaded areas correspond to the standard deviation of the metric across the whole testing dataset.}
\label{fig:rT}
\end{figure}

Next, we quantify the performance of the model using two-point statistics.
The cross-correlation and transfer function are shown in Fig.~\ref{fig:rT}.
There, the baseline linear reconstruction via Wiener filtering is actually optimistic since it uses the full 3-dimensional velocities.
In a realistic setup with only the line-of-sight component available, the quality of the linear reconstruction would be even worse.
Our U-Net improves substantially on the linear reconstruction both in cross-correlation and transfer function.
Even better results are obtained when the sub-grid and vorticity information is included with the DeepSets.
The improvement from DeepSets occurs for scales smaller than $k\sim0.1-0.2\,h\text{Mpc}^{-1}$ depending on whether cross-correlation or transfer function are considered.

\begin{figure}
\centering
\includegraphics[width=0.8\textwidth]{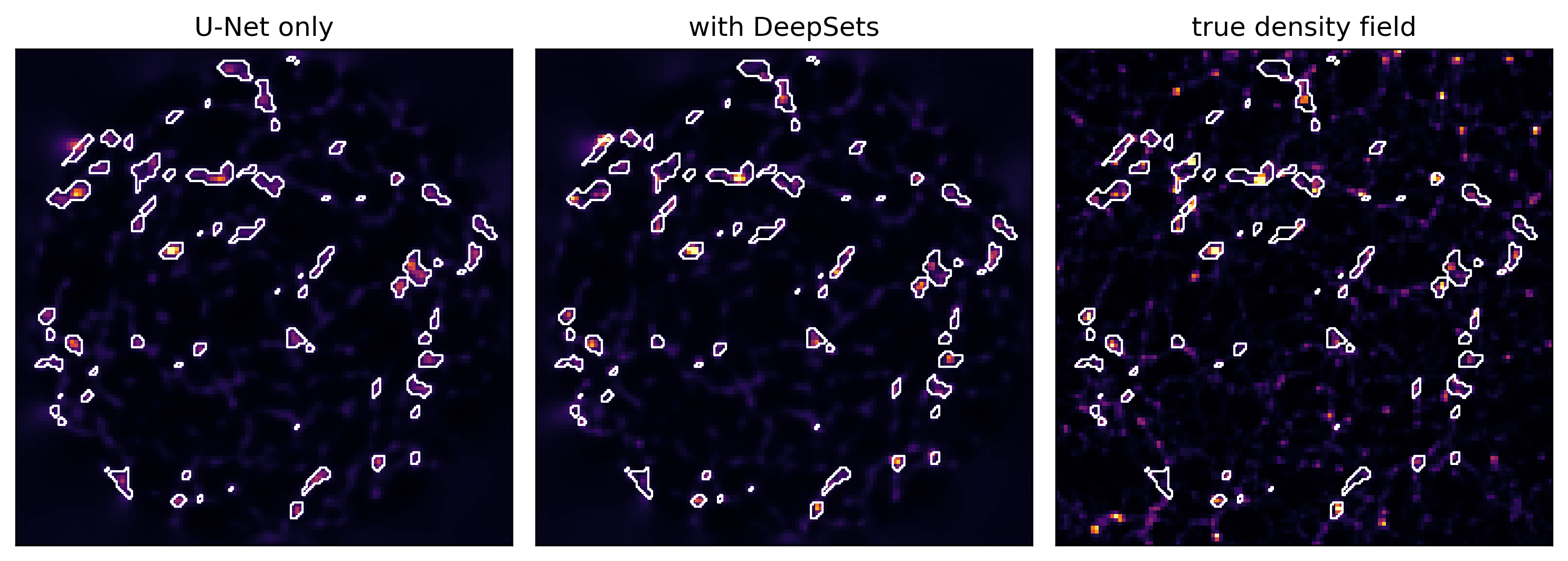}
\caption{Visual illustration of the effect of the small-scale DeepSets information. Compared to Fig.~\ref{fig:pic1}, the color scale concentrates on the high-density regime. The contours represent approximately the regions which the confidence network picked for DeepSets evaluation.}
\label{fig:pic2}
\end{figure}

In order to gain more intuition about the action of the DeepSets, we show an example in Fig.~\ref{fig:pic2}.
There, the white contours approximately delineate the regions where the confidence network returned the highest uncertainty and which are thus passed on for DeepSets evaluation.
We observe that the confidence network predominantly picks out the high-density regions.
The correlation coefficient between $\sigma$ and the underlying $\delta$ is greater than $0.8$ beyond $k=1\,h\text{Mpc}^{-1}$.
This is physically expected and implies that the DeepSets with fixed receptive field have a good chance to improve the prediction thanks to higher tracer density.
Within the regions impacted by the DeepSets, the prediction of high-density small-scale features is improved.
However, some density peaks were completely missed by the confidence network, implying that there could be room for improvement.

\section{Conclusions}

We have presented a hybrid convolutional and point-cloud neural network trained to recover the local density field from peculiar velocities.
The machine learning reconstruction dramatically improves upon linear Wiener filtering.
Our results indicate that even with moderate tracer density (below Cosmicflows-4 density) the extra small-scale information recovered by the point-cloud network is significant.
This result could change once observational errors are folded in, which is the subject of future work.
With multiple surveys gathering data on peculiar velocities, the future seems bright for machine learning solutions to the problem.
The combination of a U-Net with a confidence network-controlled DeepSets could find application in other problems in cosmology and beyond.

\section*{Acknowledgements}
BBG thanks Ben Horowitz for helpful discussions on the Wiener filter implementation.
BBG thanks the International Laboratory for Astrophysics, Neutrino and Cosmology Experiments (ILANCE) to organize research internships for Master students at the Center for Data-driven Discovery (CD3), Kavli IPMU (WPI), where this work was conducted. NMN acknowledges support from the Japan Foundation for Promotion of
Astronomy Research Grant and the Japan Foundation for Promotion of Science (JSPS) KAKENHI Grant Number 25K23373.
LT is supported by JSPS KAKENHI Grant 24K22878.

\bibliography{main}
\bibliographystyle{unsrt}

\end{document}